\documentclass[aps,prl,twocolumn,superscriptaddress,floatfix]{revtex4}
\usepackage{amsmath}
\usepackage{amsfonts}
\usepackage{amssymb}
\usepackage{graphicx}
\usepackage{cancel}
\usepackage[switch,columnwise]{lineno}
\usepackage{titlesec}
\usepackage{xcolor}
\usepackage{hyperref}
\usepackage{caption}
\makeatletter
\renewcommand{\section}{\@startsection{section}{1}{0mm}
  {-\baselineskip}{0.5\baselineskip}{\bf\leftline}}

\renewcommand{\subsection}{\@startsection{section}{1}{0mm}
  {-\baselineskip}{0.5\baselineskip}{\bf\leftline}}
\makeatother

\begin{document}
\title{Continuous variable quantum communication with 40 pairs of entangled sideband modes}

\author{Xuan Liu}%
\thanks{These authors contributed equally to this work}
\affiliation{State Key Laboratory of Quantum Optics and Quantum Optics Devices, Institute of Opto-Electronics, Shanxi University, Taiyuan 030006, China}%
\author{Shaoping Shi}%
\thanks{These authors contributed equally to this work}
\email{ssp4208@sxu.edu.cn}
\affiliation{State Key Laboratory of Quantum Optics and Quantum Optics Devices, Institute of Opto-Electronics, Shanxi University, Taiyuan 030006, China}%
\author{Yimiao Wu}%
\affiliation{State Key Laboratory of Quantum Optics and Quantum Optics Devices, Institute of Opto-Electronics, Shanxi University, Taiyuan 030006, China}%
\author{Xuan Wang}%
\affiliation{State Key Laboratory of Quantum Optics and Quantum Optics Devices, Institute of Opto-Electronics, Shanxi University, Taiyuan 030006, China}%
\author{Long Tian}%
\affiliation{State Key Laboratory of Quantum Optics and Quantum Optics Devices, Institute of Opto-Electronics, Shanxi University, Taiyuan 030006, China}%
\affiliation{Collaborative Innovation Center of Extreme Optics, Shanxi University, Taiyuan 030006, China}%
\author{Wei Li}%
\affiliation{State Key Laboratory of Quantum Optics and Quantum Optics Devices, Institute of Opto-Electronics, Shanxi University, Taiyuan 030006, China}%
\affiliation{Collaborative Innovation Center of Extreme Optics, Shanxi University, Taiyuan 030006, China}%
\author{Yajun Wang}%
\affiliation{State Key Laboratory of Quantum Optics and Quantum Optics Devices, Institute of Opto-Electronics, Shanxi University, Taiyuan 030006, China}%
\affiliation{Collaborative Innovation Center of Extreme Optics, Shanxi University, Taiyuan 030006, China}%
\author{Yaohui Zheng}%
\email{yhzheng@sxu.edu.cn}
\affiliation{State Key Laboratory of Quantum Optics and Quantum Optics Devices, Institute of Opto-Electronics, Shanxi University, Taiyuan 030006, China}%
\affiliation{Collaborative Innovation Center of Extreme Optics, Shanxi University, Taiyuan 030006, China}%

\begin{abstract}
Constructing large-scale quantum resources is an important foundation for further improving the efficiency and scalability of quantum communication. Here, we present an efficient extraction and stable control scheme of 40 pairs of entangled sideband modes from the squeezed light by specially designing optical parametric oscillator. Utilizing the low-loss optical frequency comb control technology and the local cross-correlation algorithm, we model and manage the efficient separation process of the entangled sidebands modes facilitated by the optical filtering cavities, a maximum entanglement level of 6.5 dB is achieved. The feasibility of large-capacity quantum dense coding based on these entangled sideband modes is proved experimentally, which is of great significance for optimizing the utilization of quantum resources, thereby contributing to the advancement of large-capacity quantum communication networks and enabling the realization of more secure and efficient quantum communication systems.
\end{abstract}
\maketitle

\begin{figure*}
\centering
\includegraphics[width=\textwidth]{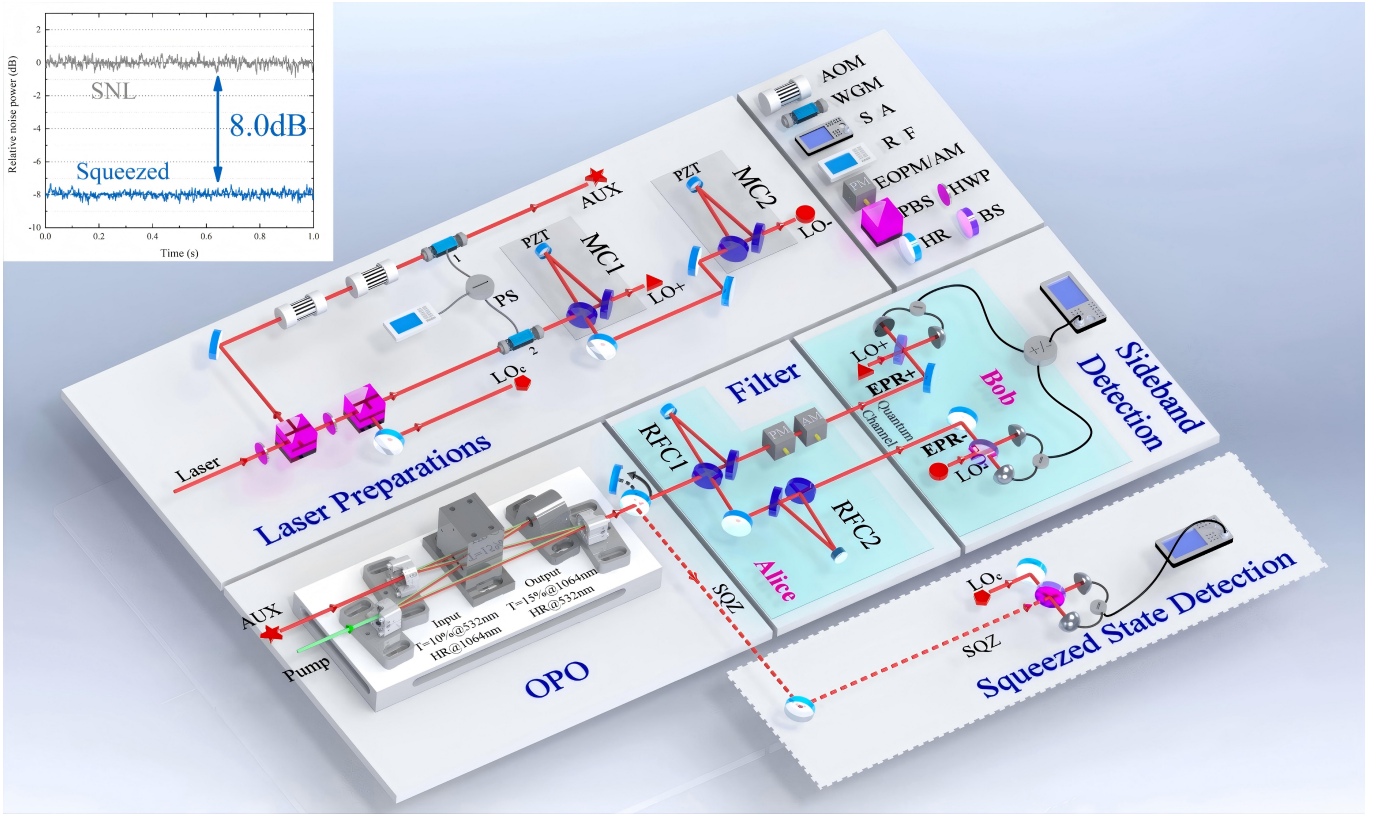}
\caption{\textbf{\label{fig1} Experimental setup for constructing a large-scale QDC system on the basis of entangled sideband modes in a free-space quantum channel.} AOM, acousto-optic modulator; WGM, waveguide electro-optic phase modulator; SA, spectrum analyzer; RF, radio frequency source; EOAM, electro-optic amplitude modulator; EOPM, electro-optic phase modulator; PBS, polarization beam splitter; HWP, half-wave plate; HR, mirror with high reflectivity; BS, beam splitter; MC, mode cleaner; RFC, ring filter cavity; SQZ, squeezed state; LO, local oscillator; OPO, optical parametric oscillator; PS, power splitter; PZT, piezoelectric transducer.}
\label{fig1}
\end{figure*}

Development in quantum technologies become increasingly important for many applications in quantum computation, quantum sensing, and quantum communication \cite{ref1,ref2,ref3,ref4,ref5}. In case of quantum communication, quantum technology has driven communication system to higher security and larger information capacity based on the nature of quantum mechanics, comparing with classical communication
\cite{ref6,ref7,ref8,ref9,ref10}. Various quantum communication protocols, including quantum teleportation, quantum key distribution, quantum secure direct communication, and quantum dense coding (QDC), have been proposed and even developed towards practical applications \cite{ref11,ref12,ref13,ref14,ref15,ref16,ref17,ref18,ref19,ref20,ref21,ref22,ref23}. QDC, as a unique quantum communication protocol, enables a single qumode to transmit two bits of classical information within continuous variable systems
\cite{ref24,ref25,ref26,ref27,ref28,ref29,ref30}.
This effectively doubles the channel capacity and optimizes the utilization of quantum resources. Currently, it has been experimentally validated that multiplexed orbital angular momentum (OAM) superposition modes and an on-chip six-user fully connected network can be employed for communication within a four-wave mixing system \cite{ref31,ref32}. The frequency division multiplexed radio-frequency-over-light (RFOL) communication within a broadband entangled source \cite{ref33}, or the channel multiplexing quantum communication based on the single squeezing source \cite{ref34}, also have been demonstrated experimentally. With the development of quantum technology, point-to-point communication protocols have been extended to multiparty networks involving multiple senders and multiple receivers
\cite{ref35,ref36,ref37,ref38}.
To meet the growing network demands and accommodate multi-user operations, the manipulation, extraction and detection of multi-channel quantum resources have become one of the key technologies urgently in need of solution. Despite the remarkable advancement has been accomplished by large-capacity quantum communication, there remains substantial potential for increasing the number of quantum channels in the context of constructing quantum communication networks
\cite{ref39,ref40,ref41,ref42}.

To construct large-scale quantum systems, the relatively low utilization rate of quantum resources has been a significant limitation to further development
\cite{ref43,ref44,ref45,ref46}.
Via the concept of time division multiplexing, large-scale one-dimensional and two-dimensional continuous variable cluster states were constructed in which the modes are detected in turn, with only a few modes accessible at finite time
\cite{ref47,ref48}.
Instead with time-multiplexing protocol, 15 pairs of quadripartite entangled states with 4 dB, and multipartite entanglement of 60 adjacent modes with 3.2 dB over 60 modes of quantum optical frequency comb were generated from a dual-polarized optical parametric oscillator (OPO), respectively
\cite{ref49,ref50}.
In addition, 10-mode entanglement state was observed from the parametric downconversion of femtosecond-frequency combs \cite{ref51}. The number of entangled modes was limited by measurement technique that cannot juggle the bandwidth and efficiency. By the employment of unbalanced homodyne detection scheme, 18-teeth of a squeezing comb was simultaneously detected with a quantum noise reduction of 9.3 dB \cite{ref52}. Recently, continuous variable multipartite entanglement was also generated on an integrated optical chip with 2 dB, indicating the possibility of scaling photonic quantum technologies \cite{ref53}. These above-mentioned protocols can enable scalable and universal large-scale quantum computation. However, the upper and lower entangled sideband modes of entanglement state were evaluated by bichromatic detection, not spatially separated, limiting its application in some scenes, such as multiple-users quantum communication \cite{ref34}. It is urgent to develop a management scheme of separating the upper and lower entangled sideband modes with enough mode number, meeting the requirement for multiple-user quantum communication \cite{ref54,ref55,ref56,ref57}.

In this manuscript, we successfully generate 40 pairs of entangled sideband modes over 84 modes of the quantum optical frequency comb, with a maximum entanglement level of 6.5 dB, demonstrating channel multiplexing QDC communication. The maximum frequency shift of the entangled modes from carrier is limited by the bandwidth (20 GHz) of the waveguide electro-optic phase modulator (WGM). We build an OPO with a free spectral range (FSR) of 474.7 MHz to reduce the frequency interval between two neighboring channels. However, the small channel interval makes the separation of the upper and lower entangled sideband modes face challenge. By employing the local cross-correlation (LCC) algorithm, we model the efficient separation process of the entangled sidebands modes facilitated by the optical filtering cavities, generating the maximum number of entangled sideband modes from a single quantum resource of squeezed state to my knowledge. Each quantum channel has large channel capacities exceeding the Holevo limit of single-mode bosonic channels (Fock state) within large average photon number \cite{ref20,ref58}. Our approach significantly raises the amount of available entanglement channels, providing new insights for the development of multi-user continuous-variable quantum communication networks.

\begin{figure*}
\centering
\includegraphics[width=\textwidth]{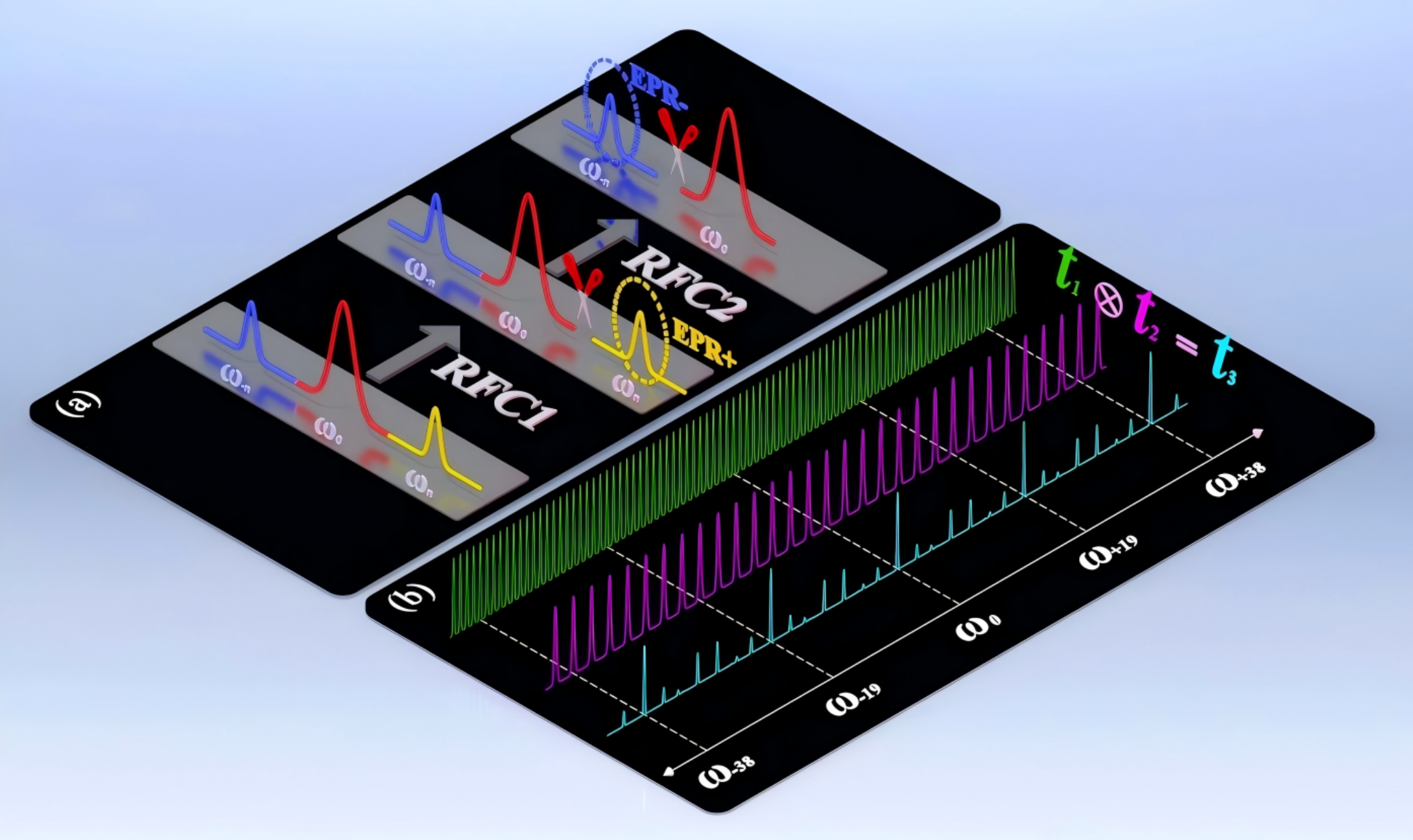}
\caption{\textbf{\label{fig2} Schematic diagram of the spatial separation of entangled sideband modes.} (a) Two cascade RFCs are adopted to transmit the $\omega_{+n}$ (EPR+) and $\omega_{-n}$ (EPR-),  respectively. (b) Overlap degree between the OPO and RFCs for entangled sideband modes of different orders. $t_1$, transfer function of the OPO; $t_2$, transfer function of the RFCs; $t_3$, overlap degree between the $t_1$ and $t_2$; $\omega_{0}$, squeezing carrier; $\omega_{i}$, entangled sideband modes ($i = 1,\ 2,\ 3,\ \cdots42)$. Since the RFCs and MCs are identical, this figure can also represent the separation of the modulated upper and lower sidebands by MCs to prepare LOs.}
\end{figure*}

\begin{figure*}[ht]
\centering
\includegraphics[width=\textwidth]{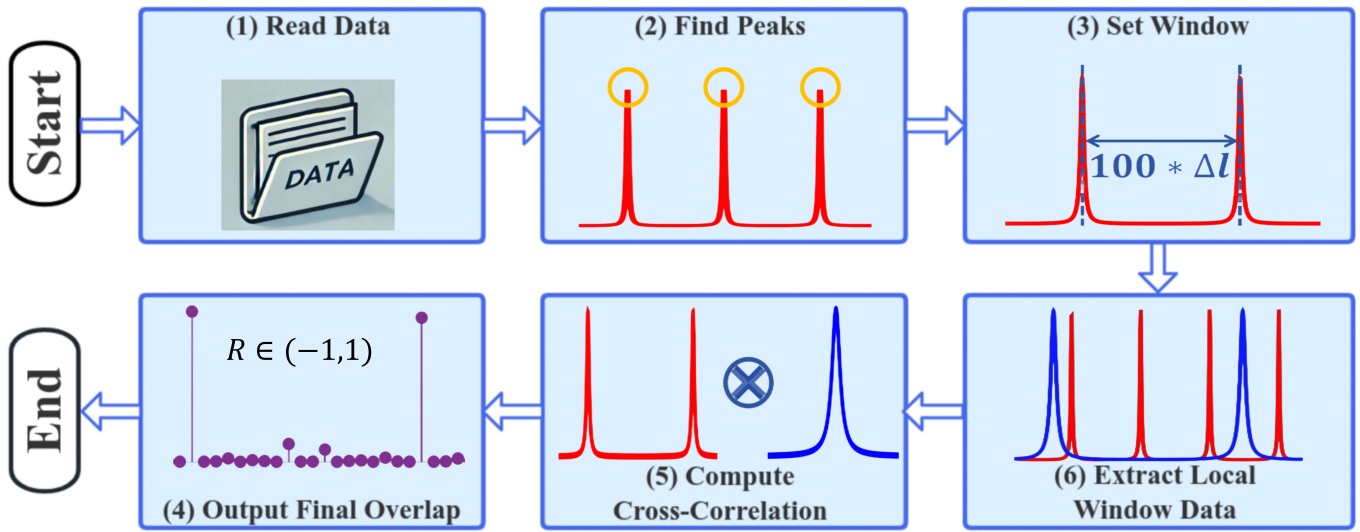}
\caption{\textbf{\label{fig3} Main flowchart of the proposed algorithm for the LCC.}}
\end{figure*}

\section*{\textbf{Experimental Setup}}
Figure~\ref{fig1} shows the experimental setup for constructing a large-scale quantum communication system via entangled sideband modes. Squeezed state of light is generated by a below-threshold OPO operating at deamplification state, the output of which is spatially separated to construct the entangled sideband modes. It mainly includes four parts: the squeezed state preparation included OPO, the preparation for local oscillator (LO) and auxiliary beams (AUX), the filter module to separate sideband modes, and the balanced homodyne detector (BHD) joint measurement. In addition, the dotted plate serves as an extra part that is used to detect the quantum noise of the squeezed carrier produced by the OPO.

The OPO cavity consists of two concave mirrors, two convex mirrors, and one 1.0 mm $\times$ 2.0 mm $\times$ 10.5 mm periodically poled $\mathrm{KTiOPO_4}$ (PPKTP) crystal, resulting in the FSR of 474.7 MHz and linewidth of 12.36 MHz. The input and output mirrors are designed as convex mirrors to minimize the effect of astigmatism \cite{ref59}. The input coupling mirror has a transmissivity of 10\% for the harmonic wave and high reflectivity (HR) for the fundamental beam. The output coupling mirror has a transmissivity of 15\% for the fundamental beam and HR for the harmonic wave. The two concave mirrors are coated as HR for two wavelengths of 1064 nm and 532 nm. The pump light power of 55 mW is produced by a second harmonic generator that has a same design with Ref. \cite{ref60}. The squeezing angle is controlled by sensing the frequency-shifted field AUX with coherent amplitudes \cite{ref61}. By finely tuning the crystal temperature, the fundamental beam and the pump light are ensured to resonate simultaneously within the cavity. Moreover, the pump light is utilized to control the OPO cavity length, guaranteeing it works in the status of maximum squeezed for optimal output. The corresponding data plotted in the upper left corner of Fig. 1 shows that the squeezed carrier has a quantum noise reduction of 8.0 $\pm$ 0.1 dB at the analysis frequency of 500 kHz. According to the escape efficiency of the OPO (97.8 $\pm$ 0.5\%), the interference visibility (96.0 $\pm$ 0.1\%), the propagation efficiency (97.2 $\pm$ 0.2\%) and the quantum efficiency of the photodiode (98.5 $\pm$ 0.1\%), the total efficiency is calculated to be 86.3 $\pm$ 0.2\%.

A squeezed field involves many Einstein-Podolsky-Rosen (EPR) entangled sideband modes that is symmetric distribution around the half pump frequency. Any two neighboring modes have equal frequency interval that is the FSR of the OPO cavity. Each pair of entangled sideband modes can serve as independent quantum channel for QDC communication. In order to spatially separate the upper and lower sideband modes, we employ two ring filter cavities (RFCs) with optimal impedance matching and high sideband suppression ratio  \cite{ref62}. However, these sideband modes have no coherent amplitudes, we cannot directly extract the error signal of the downstream experiment. To address this issue, a fiber-coupled WGM1 is employed to generate a frequency comb-like AUX. In previous works, the AUX was phase-locked to the squeezing carrier to manage the cavities and relative phase in downstream experiments \cite{ref34}. Unfortunately, the above-mentioned scheme requires an additional phase-locked loop, which inevitably introduces extra loss, leading to a reduction in the squeezing factor. Therefore, we employ the AUX as the seed beam of the OPO, avoiding the additional loss in the coupling process \cite{ref63}. In addition, to reduce the influence of the technical noise at low frequencies on noise variances, we utilize two additional acousto-optic modulators (AOMs) that are placed in front of the WGM1 to generate a frequency shift of 12 MHz from the corresponding vacuum sideband modes of squeezed field, with a diffraction efficiency of 70\% for each AOM. The frequency comb-like AUX with coherent amplitudes serves as the reference beam for the error signal extraction in the downstream experiment, without inducing the nonlinear noise coupling \cite{ref64}, extending the entangled sideband modes to kilohertz frequency band. 

Considering the phase-matching bandwidth of the PPKTP is on the terahertz magnitude, the frequency shift range of entangled sideband modes is limited by the modulation bandwidth of the WGMs. Therefore, within the bandwidth of the WGMs, we attribute the key factors that restricting the number of the detectable entangled sideband modes and the level of entanglement to the design of the RFC1 and RFC2. The detailed principle of the frequency-dependent beam splitters RFC1 and RFC2 in the filter section is shown in Figure~\ref{fig2}(a). It follows that the RFC1 resonates with the upper sideband and transmits it for acting as the EPR+, meanwhile reflecting the rest sidebands. Then the RFC2 further transmits the lower sideband for acting as the EPR-. The transfer function $t(f)$ of the frequency comb-like sideband modes of the optical resonant cavities can be expressed as:
\begin{equation}
t(f)=\frac{\sqrt{(1-r_1^2)(1-r_2^2)}\ e^{\frac{i\pi f}{\omega_{FSR}}}}{1-r_1r_2\ e^\frac{2i\pi f}{\omega_{FSR}}},
\end{equation}
where $r_i$ denotes the reflectance of the input and output mirrors, $\omega_{FSR}$ is the FSR of the cavity, and $f$ represents the Fourier frequency. Since the separation efficiency of entangled sideband modes is directly limited by the relative FSR of the RFCs and OPO, thus the lengths of the RFCs must be fine-tuned. We proposed an alternative LCC algorithm, which aims to achieve the visualized analysis of the overlap degree of sidebands. The detailed steps of the LCC algorithm are outlined in Figure~\ref{fig3}, as shown below.

\begin{figure*}
\centering
\includegraphics[width=\textwidth]{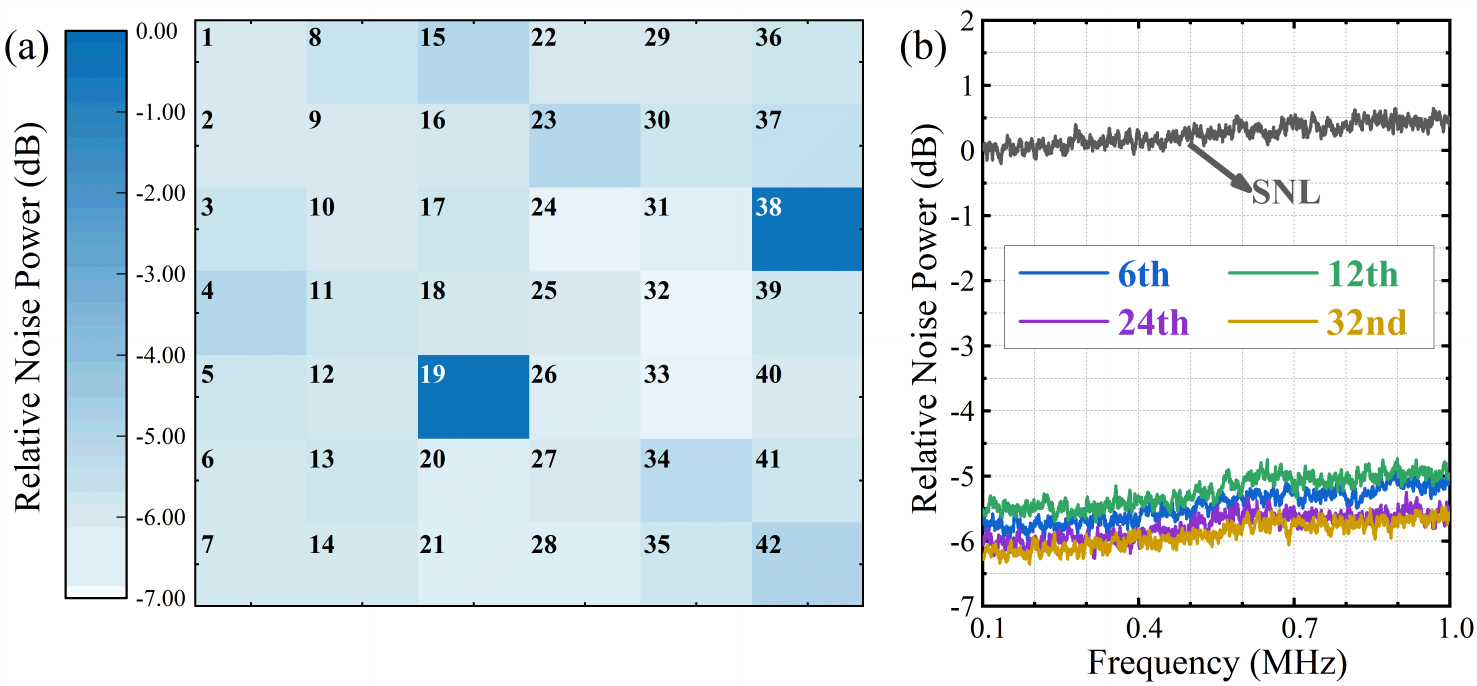}
\caption{\textbf{\label{fig4} Quantum noise levels of the entangled sideband modes.} (a) Relative noise power for unbiased entangled sideband modes of orders 1 to 42 measured at the analysis frequency of 500 kHz. (b) Relative noise power at the $6^{th}$, $12^{th}$, $24^{th}$ and $32^{nd}$ orders of unbiased entangled sideband modes measured at the analysis frequencies of 0.1 - 1 MHz. The resolution bandwidth (RBW) is 20 kHz and the video bandwidth (VBW) is 50 Hz. SNL, shot noise limit.}
\end{figure*}

\begin{enumerate}
    \item[1)]Discretize the transmission plots of both the OPO and RFCs, and extract the corresponding frequency and amplitude values to construct the datasets. The frequency range from 0 to 40 GHz are considered as the independent variable, with a step of 1 kHz.
\end{enumerate}
\begin{enumerate}
    \item[2)]These peaks in the amplitude datasets of both the OPO and RFCs are identified and annotated
\end{enumerate}
\begin{enumerate}
	\item[3)]Since the FSR of the OPO is smaller in value compared to that of the RFCs, the window width is set to 1\% of the OPO's FSR, which corresponds to 4.747 MHz. The range of window width is between 0.2 and 1 times of the  OPO's linewidth.
\end{enumerate}
\begin{enumerate}
  	\item[4)]For each window of the OPO, the corresponding local window in the RFC data is identified, and then both are extracted and interpolated to the same frequency range.
\end{enumerate}
\begin{enumerate}
	\item[5)]The overlap degree between the OPO and RFCs is determined by calculating their cross-correlation coefficient. Let $X$ and $Y$ denote the sets of amplitude data for OPO and RFCs, respectively, where $X_i\in X$ and $Y_i\in Y$ ($i=1,\ 2,\ 3,\ \cdots$) represent individual elements of the corresponding sets. The overlap between them is evaluated by calculating the Pearson correlation coefficient $R$ within the given window, as given by the following formula \cite{ref65}:
	\begin{equation}
		R=\frac{\sum_{i=1}^{n}\left(X_iY_i\right)}{\sqrt{\frac{1}{n-1}\sum_{i=1}^{n}\left(X_i\right)^2}\sqrt{\frac{1}{n-1}\sum_{i=1}^{n}\left(Y_i\right)^2}}.
	\end{equation}
	The correlation coefficient $R\in\left(-1,1\right)$, where the strength of the linear relationship between datasets $X$ and $Y$ increases as the absolute value of $R$ approaches 1.
\end{enumerate}
\begin{enumerate}
	\item[6)]Record the maximum cross-correlation coefficient for each peak, and output the resulting overlaps.
\end{enumerate}

We utilized the LCC algorithm to fit the sideband separating ability of the RFCs under the distinct FSR. As shown in Figure~\ref{fig2}(b), where the light blue line represents the overlap degree between the transfer functions of the OPO and RFCs. Since the frequency interval between upper and lower sidebands is $2n$ times the FSR of the OPO, the 84 spectral teeth are presented within the twice modulation bandwidth of the WGM (2 $\times$ 20 GHz). By normalizing the transfer functions of the OPO and RFCs, we can observe the overlap degree more intuitively, so as to determine the optimal RFCs parameters. Following iterative optimization, the FSR of the two RFCs was set to 1.288 GHz, achieving a maximum of 40 pairs of separable entangled sideband modes. It is evident that there is a significant overlap between the upper and lower sidebands of the 19th and 38th orders entangled sideband modes, preventing the separation of them. We abandon these two orders of entangled sideband modes to ensure the separation efficiency of the remaining sidebands. In additional, we control the cavity length of the RFCs by employing the same technique as the Ref. \cite{ref66}. Both RFCs consist of two plane mirrors with transmissivity of $T=10\%$ and one concave HR mirror with a curvature of $r=-1000$. The concave HR mirror is bonded to piezoelectric transducer for controlling the cavity length. Furthermore, the linewidth of the RFCs is 43.4 MHz, which is broader than that of the OPO, ensuring that the entangled sideband modes are effectively separated without reducing their bandwidth \cite{ref34}.

The separated entangled sideband modes are interfered with the corresponding LOs on a 50/50 beam splitter and directed toward two BHDs to detect the correlation noise. The generation process of the single-sideband LOs (LO+ and LO-) is identical to that of the EPR+ and the EPR-. The mode cleaners (MCs) consist of two plane mirrors with transmissivity of $T=1\%$ and one concave HR mirror with a curvature of $r=-1000$. The finesse value of up to 312 ensures that we can measure the quantum correlation variance of the required sideband modes while effectively avoiding interference from the neighboring resonant modes.

To meet the power requirements of the various beams involved in the discussed earlier, the primary laser source is split into several parts. Approximately 120 mW of the laser is picked off for the generation of the AUX beam. In order to prepare two teeth of the single sideband LO that is indispensable for implementing BHDs joint measurement during the detection of entangled sideband modes, 20 mW of power is injected to the two cascaded MCs from the rear of the WGM2. Both WGM1 and WGM2 have optical transmittance of 40\% and are driven by a single radio frequency source [E8257D, Keysight], which has a bandwidth and output power of 20 GHz and 25 dBm, respectively.

Ultimately, the quantum correlation noise of 42 pairs of entangled sideband modes measured experimentally at the analysis frequency of 500 kHz are shown in Figure~\ref{fig4}(a). The number in each square represents the order of the entangled sideband modes, meanwhile the shade of color represents the relative noise power. By switching the frequency of the modulation of the WGM1 and WGM2 to the various of entangled sideband mode orders, 40 pairs of entangled sideband modes are experimentally observed, with a maximum entanglement level of 6.5 dB. Compared to the squeezed carrier detected directly from the OPO, the additional RFCs introduce the inevitable loss in the optical path, hence resulting in the degradation of quantum correlation. The optical loss introduced by RFCs is primarily attributed to the mode-matching efficiency, the impedance-matching, and the light scattering effect, which collectively determine the lower bound of optical losses. On the other hand, there is a slight variation in the entanglement levels between the 40 pairs of entangled sideband modes measured. These differences mainly attribute to the partial overlap between the OPO and RFCs, the measured entanglement level degrades as the overlap degree increases. Each pairs of entangled sideband modes serve as the fundamental physical resource for quantum communication. By the employment of the entangled sideband modes generated here, we expect to demonstrate channel multiplexing quantum communication and quantum network with various topologies \cite{ref67}. Here, we make an exemplary choice of four pairs of entangled sideband modes with different entanglement levels for the QDC demonstration. The Figure~\ref{fig4}(b) shows the relative noise power of selected four pairs of entangled sideband modes within the 0.1-1 MHz frequency range. Since the variances of the amplitude sum and phase difference are unbiased, only the relative noise power of the amplitude sum is shown. 

\section*{\textbf{Experimental Results}}
\begin{figure}
\centering
\includegraphics[width=0.48\textwidth]{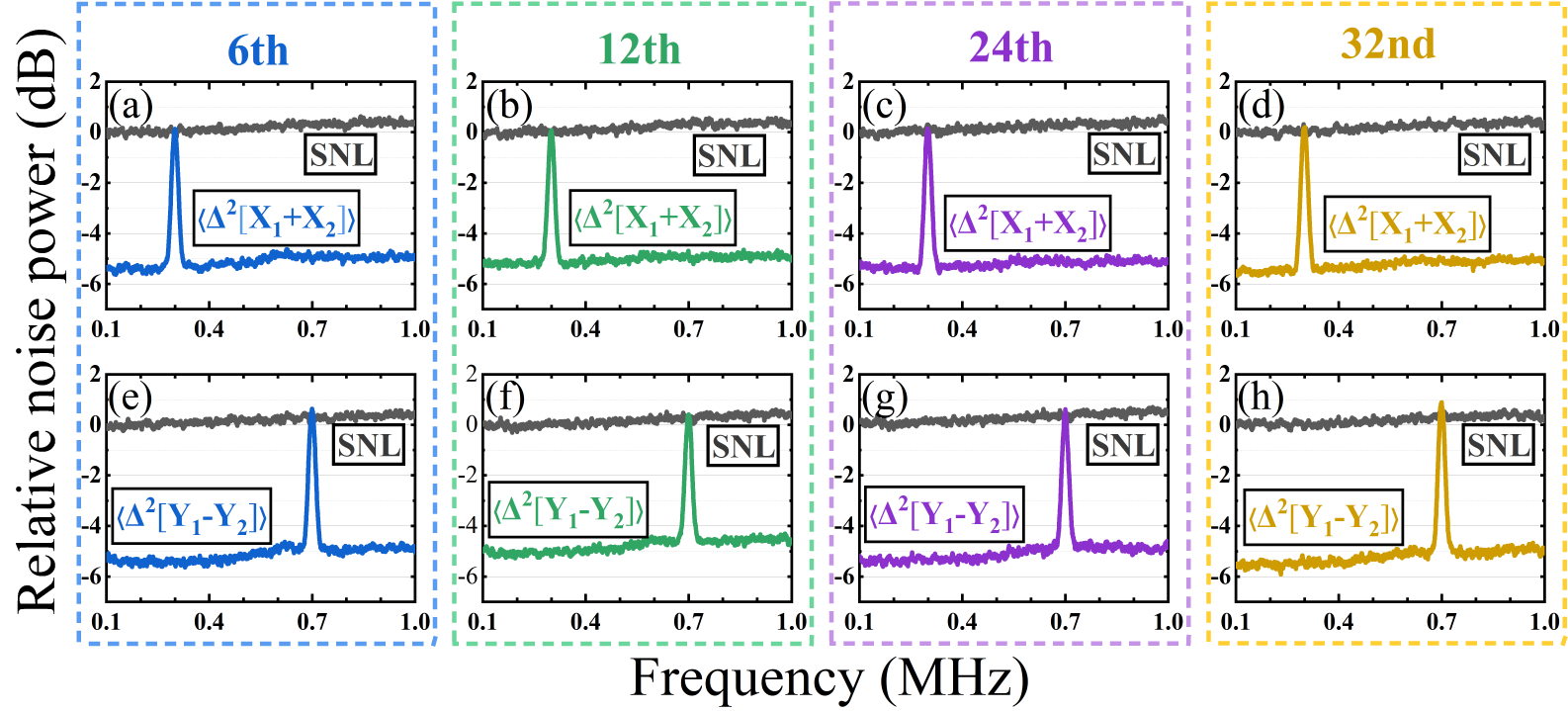}
\caption{\textbf{\label{fig5} Spectral densities of amplitude sum $\langle\Delta^{2}[X_1+X_2]\rangle$ (a)-(d) and phase difference $\langle\Delta^{2}[Y_1-Y_2]\rangle$ (e)-(h) at the $6^{th}$, $12^{th}$, $24^{th}$ and $32^{nd}$ orders of entangled sideband modes.} Multiple encoding signals operating at 300 kHz and 700 kHz surpass the SNL exceeds 5 dB through the utilization of symmetric entangled sidebands. RBW, 20 kHz; VBW, 50 Hz.}
\end{figure}

Leveraging the multi-order unbiased entangled sideband modes, the feasibility of QDC communication was verified experimentally. The results of encoding and decoding $\omega_{\pm6}$, $\omega_{\pm12}$, $\omega_{\pm24}$ and $\omega_{\pm32}$ (corresponding to the $6^{th}$, $12^{th}$, $24^{th}$ and $32^{nd}$ EPR entangled sideband modes) are shown in Figure~\ref{fig5}. Alice encodes one of the submodes in the entangled sideband modes, and the information from the other submode is shared\hfill
\noindent between Alice and Bob. Then Alice transmits the encoded submode (EPR+) to Bob, who decodes the received information with the assistance of the other shared submode (EPR-). It is obvious that the extraction of signals merged in the large noise background is quite difficult if without the help of the other half of the EPR beam. Due to the presence of EPR entanglement, the signal-to-noise ratio (SNR) of the signals decoded in this way can surpass the shot noise limit (SNL), enabling higher channel capacity than classical communications. Notably, individual submode always as a thermal state during the measurement, thus enhancing the security of communication. During the experiment, we employed electro-optic amplitude modulator (EOAM) and electro-optic phase modulator (EOPM) to encode the amplitude and phase quadratures, respectively. This process inevitably introduces additional losses, leading to a reduction in the measured entanglement level compared to Figure~\ref{fig4}(b). The information decoded at 300 kHz in the amplitude and phase quadratures within the four pairs of entanglement sideband modes described above, supported by the shared entanglement submodes, exhibits the enhancements in SNR of 5.4, 5.1, 5.3, and 5.5 dB relative to the classical communication system. 

In the large average photon number ($\bar{n}$) regime, if the channel capacity of the QDC scheme is greater than the ultimate single-mode communication limit of the Fock state with non-Gaussian operations, it can be considered that the unconditional dense coding has been achieved. Therefore, the actual capacity of a single QDC channel is accurately compared with the Fock state communication limit from the perspective of channel capacity \cite{ref30,ref34,ref68}. After subtracting the influence of photodetection efficiency (including interference visibility and photodiode quantum efficiency) and electronic noise, the actual noise variances in phase and amplitude quadratures of the entangled sideband modes at 300 kHz
\begin{figure}
\centering
\includegraphics[width=0.48\textwidth]{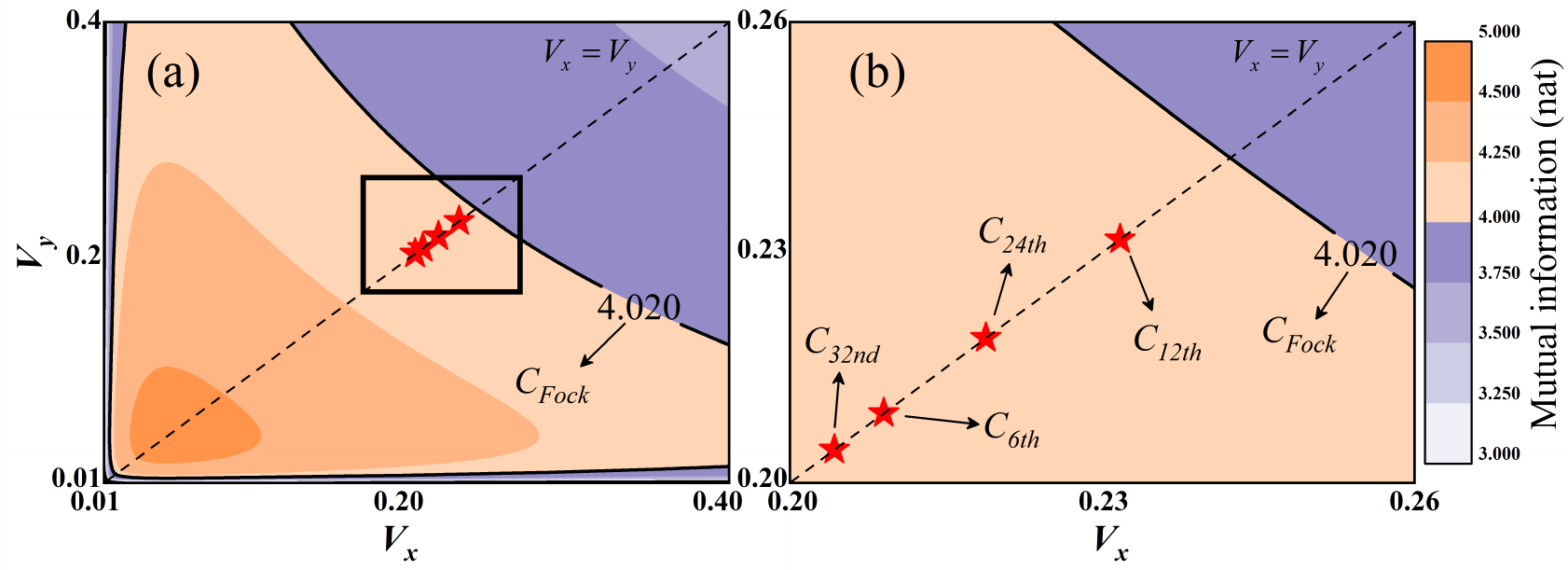}
\caption{\textbf{\label{fig6} Contour map illustrates the dependence of the mutual information on the quantum noise variance in $V_x$ and $V_y$ with the average photon number $\bar{n}=20$.} (a) Channel capacity of the $6^{th}$, $12^{th}$, $24^{th}$ and $32^{nd}$ orders of entangled sideband modes. (b) Zoom in of the state. The dashed lines indicate the scenarios that the $V_y$ is equal to the $V_x$. $C_{Fock}$, Holevo limit of a single-mode bosonic channel (When $\bar{n}=20$, $C_{Fock}=4.020$); $C_{6th}$, $C_{12th}$, $C_{24th}$, $C_{32nd}$, channel capacity of different-order entangled sideband modes.}
\end{figure}
\noindent are $V_{x\left(y\right),6th}=0.209$, $V_{x\left(y\right),12th}=0.232$, $V_{x\left(y\right),24th}=0.219$ and $V_{x\left(y\right),32nd}=0.204$, respectively. For signal (S) and noise (N) with a Gaussian distribution, the channel capacity $C$ can be given by:
\begin{equation}
C=\frac{1}{2}\ln{\left[1+\frac{S}{N}\right]}.
\end{equation}
Accordingly, the comparison between the channel capacity of a single QDC communication and that of the Fock state communication is shown in Figure~\ref{fig6}. Evidently, for the four channels shown in the figure, the channel capacity of QDC communication exceeds the Holevo limit of a single-mode bosonic channel, as well as that of other CV quantum communication protocols, including channels such as squeezed state at their optimal values. Furthermore, we indicate that our experimental system can meet the requirements for implementing unconditional QDC if the average photon number exceeds 20 and the SNR below the SNL is still more than 4.9 dB.

During the experiment, we propose that the RFC2 can be completely omitted, which further reduces the complexity of implementing channel multiplexing quantum communication technology in practical applications. Nevertheless, utilizing only a single RFC to separate the upper and lower sidebands will introduces additional phase delay, leading to bandwidth limitation of generating the sideband entangled modes. The additional phase delay in the transmitted field is necessary, and the retardation is dependent of the RFC's linewidth. By introducing a corresponding electronic delay at BHD2's output based on the RFC's phase delay, we effectively compensate the phase shift while preserving the full entanglement bandwidth  \cite{ref69}. By leveraging the currently detected multi-order entangled sideband modes and designing different optical filter cavities, precise control over each sideband can be achieved. 
	
Compared to Ref. \cite{ref31,ref49,ref50,ref53,ref70}, our protocol achieves both effective spatial separation of entangled sideband modes and remarkable performance in terms of both the quantity and quality of detected entanglement, fully supporting the implementation of various quantum communication protocols, including quantum conference key agreement \cite{ref71}, entanglement swapping-based quantum relays \cite{ref72} and multi-node quantum communication networks \cite{ref73,ref74}. As for variable multiplexing techniques such as spatial, polarization, or time-division multiplexing, frequency-division multiplexing exhibits inherent advantages in crosstalk suppression and classical communication compatibility. These distinctive characteristics make it particularly suitable for high-capacity information transmission systems. We expect to further increase the number of sideband modes by employing wider-bandwidth WGMs, cascaded WGMs, or parametric homodyne detection techniques. This approach offers new research perspectives for the advancement of quantum communication technologies.

\section*{\textbf{Conclusions}}

In summary, we have demonstrated the generation and detection of multiple pairs of entangled sideband modes derived from a single quantum resource of squeezed state. Based on a specially designed bow-tie OPO and creative LCC algorithm, up to 40 pairs of sideband modes are directly detected, with a maximum entanglement level of 6.5 dB. This level of entanglement ensures robust quantum information transfer even under constrained channel conditions. We also propose that channel multiplexing QDC schemes can be successfully executed using entangled sideband modes, where the capacity of each channel satisfy the conditions of exceeding the Holevo limit of single-mode bosonic channels (Fock state). By improving the escape efficiency of the OPO and finely optimizing the interference visibility, the detected level of the entanglement can be enhanced, thereby further improving the channel capacity. Our proof-of-principle experiment can realize the construction and utilization of large-capacity entanglement resources with a single squeezed field. By combining time domain multiplexing through synchronized timing signals and adjustable delay compensation \cite{ref47,ref48}, and high-dimensional entanglement via optical mode conversion \cite{ref75}, we expect to significantly increase the channel number.

\section*{\textbf{Data availability}}
All data needed to evaluate the conclusions in the paper are present in  the paper and/or the Supplementary Information. Additional data related to this paper may be requested from the corresponding author.

\section*{\textbf{Acknowledgements}}

We acknowledge the financial support from the National Natural Science Foundation of China (NSFC) (Grants No. 62225504, No. 62027821, No. 62035015, No. 12174234, No. 11874250, No. 12274275); the National Key R\&D Program of China (Grant No. 2020YFC2200402); Key R\&D Program of Shanxi (Grant No. 202102150101003); Program for Sanjin Scholar of Shanxi Province. H. Shen acknowledges the financial support from the Royal Society Newton International Fellowship Alumni follow-on funding (AL201024) of United Kingdom.

\section*{\textbf{Competing interests}}
The authors declare no competing interests.


\begin{thebibliography}{9}\label{sec:TeXbooks}%
\section*{\textbf{References}}
\bibitem {ref1} M. V. Larsen, X. S. Guo, C. R. Breum, J. S. Neergaard-Nielsen, and U. L. Andersen, Science 366, 369 (2019).

\bibitem {ref2} N. C. Menicucci, Phys. Rev. Lett. 112, 120504 (2014).

\bibitem {ref3} H.-L. Huang, X.-Y. Xu, C. Guo, G. Tian, S.-J. Wei, X. Sun, W.-S. Bao, and G.-L. Long, Sci. China-Phys. Mech. Astron. 66, 250302 (2023).

\bibitem {ref4} J. M. Boss, K. S. Cujia, J. Zopes, and C. L. Degen, Science 356, 937 (2017).

\bibitem {ref5} A. Ruskuc, C.-J. Wu, E. Green, S. L. N. Hermans, W. Pajak, J. Choi, and A. Faraon, Nature 639, 54 (2025).

\bibitem {ref6} S. Wengerowsky, S. K. Joshi, F. Steinlechner, H. H{\"u}bel, and R. Ursin, Nature 564, 225 (2018).

\bibitem {ref7} R. Qi, Z. Sun, Z. Lin, P. Niu, W. Hao, L. Song, Q. Huang, J. Gao, L. Yin, and G. L. Long, Light Sci. Appl. 8, 22 (2019).

\bibitem {ref8} C. Liu, C. Zhang, S.-P. Gu, X.-F. Wang, L. Zhou, and Y.-B. Sheng, Sci. China-Phys. Mech. Astron. 68, 250311 (2025).

\bibitem {ref9} L. S. Madsen, V. C. Usenko, M. Lassen, R. Filip, and U. L. Andersen, Nat. Commun. 3, 1083 (2012).

\bibitem {ref10} S. K. Joshi, D. Aktas, S. Wengerowsky, M. Lon{\"o}ari{\'c}, S. P. Neumann, B. Liu, T. Scheidl, G. C. Lorenzo, {\v{Z}}. Samec, L. Kling, A. Qiu, M. Razavi, M. Stip{\v{c}}evi{\'c}, J. G. Rarity, and R. Ursin, Sci. Adv. 6, 36 (2020).

\bibitem {ref11} D. Bouwmeester, J.-W. Pan, K. Mattle, M. Eibl, H. Weinfurter, and A. Zeilinger, Nature 390, 575 (1997).

\bibitem {ref12} A. Furusawa, J. L. S{\o}rensen, S. L. Braunstein, C. A. Fuchs, H. J. Kimble, and E. S. Polzik, Science 282, 706 (1998).

\bibitem {ref13} S. Pirandola, J. Eisert, C. Weedbrook, A. Furusawa, and S. L. Braunstein, Nat. Photon. 9, 641 (2015).

\bibitem {ref14} D. Main, P. Drmota, D. P. Nadlinger, E. M. Ainley, A. Agrawal, B. C. Nichol, R. Srinivas, G. Araneda, and D. M. Lucas, Nature 638, 383 (2025).

\bibitem {ref15} Y. Tian, P. Wang, J. Liu, S. Du, W. Liu, Z. Lu, X. Wang, and Y. Li, Optica 9, 492 (2022).

\bibitem {ref16} C. Zhou, X. Y. Wang, Z. G. Zhang, S. Yu, Z. Y. Chen, and H. Guo, Sci. China-Phys. Mech. Astron. 64, 260311 (2021).

\bibitem {ref17} L. Wang, G. Chai, Z. Cao, X. Chen, K. Liang, and J. Peng, Sci. China-Phys. Mech. Astron. 68, 220313 (2025).

\bibitem {ref18} Z. Cao, Y. Lu, G. Chai, H. Yu, K. Liang, and L. Wang, Research 6, 0193 (2023).

\bibitem {ref19} L. Zhou, Y.-B. Sheng, and G.-L. Long, Sci. Bull. 65, 12 (2020).

\bibitem {ref20} C. H. Bennett, and S. J. Wiesner, Phys. Rev. Lett. 69, 2881 (1992).

\bibitem {ref21} K. Mattle, H. Weinfurter, P. G. Kwiat, and A. Zeilinger, Phys. Rev. Lett. 76, 4656 (1996).

\bibitem {ref22} T. Schaetz, M. D. Barrett, D. Leibfried, J. Chiaverini, J. Britton, W. M. Itano, J. D. Jost, C. Langer, and D. J. Wineland, Phys. Rev. Lett. 93, 040505 (2004).

\bibitem {ref23} Y. Yeo, and W. K. Chua, Phys. Rev. Lett. 96, 060502 (2006).

\bibitem {ref24} J. Zhang, Phys. Rev. A 67, 054302 (2003).

\bibitem {ref25} Y. Guo, B.-H. Liu, C.-F. Li, and G.-C. Guo, Adv. Quantum Technol. 2, 1900011 (2019).

\bibitem {ref26} S. L. Braunstein, and H. J. Kimble, Phys. Rev. A 61, 042302 (2000).

\bibitem {ref27} A. Piveteau, J. Pauwels, E. H{\aa}kansson, S. Muhammad, M. Bourennane, and A. Tavakoli, Nat. Commun. 13, 7878 (2022).

\bibitem {ref28} X. Li, Q. Pan, J. Jing, J. Zhang, C. Xie, and K. Peng, Phys. Rev. Lett. 88, 047904 (2002).

\bibitem {ref29} S. Liang, J. Cheng, J. Qin, J. Li, Y. Shi, B. Zeng, Z. Yan, X. Jia, C. Xie, and K. Peng, Laser {\&} Photonics Rev. 18, 2400094 (2024).

\bibitem {ref30} J. Jing, J. Zhang, Y. Yan, F. Zhao, C. Xie, and K. Peng, Phys. Rev. Lett. 90, 167903 (2003).

\bibitem {ref31} Y. Chen, S. Liu, Y. Lou, and J. Jing, Phys. Rev. Lett. 127, 093601 (2021).

\bibitem {ref32} P. Zhu, Y. Wang, Y. Du, M. Yu, K. Zhang, K. Wang, and P. Xu, Sci. China-Phys. Mech. Astron. 68, 260311 (2025).

\bibitem {ref33} S. Liang, J. Cheng, J. Qin, J. Li, Y. Shi, Z. Yan, X. Jia, C. Xie, and K. Peng, Phys. Rev. Lett. 132, 140802 (2024).

\bibitem {ref34} S. Shi, L. Tian, Y. Wang, Y. Zheng, C. Xie, and K. Peng, Phys. Rev. Lett. 125, 070502 (2020).

\bibitem {ref35} D. Bru$\beta$, G. M. D'Ariano, M. Lewenstein, C. Macchiavello, A. Sen (De), and U. Sen, Phys. Rev. Lett. 93, 210501 (2004).

\bibitem {ref36} D. Bru$\beta$, M. Lewenstein, A. Sen (De), U. Sen, G. M. D'Ariano, and C. Macchiavello. Int. J. Quantum Inf. 4, 415 (2006).

\bibitem {ref37} T. Das, R. Prabhu, A. Sen (De), and U. Sen, Phys. Rev. A 92, 052330 (2015).

\bibitem {ref38} R. Prabhu, A. K. Pati, A. Sen (De), and U. Sen, Phys. Rev. A 87, 052319 (2013). 

\bibitem {ref39} S. Shi, Y. Wang, L. Tian, W. Li, Y. Wu, Q. Wang, Y. Zheng, and K. Peng, Laser {\&} Photonics Rev. 10, 1002 (2022).

\bibitem {ref40} Y. Ren, X. Wang, Y. Lv, D. Bacco, and J. Jing, Laser {\&} Photonics Rev. 16, 2100586 (2022).

\bibitem {ref41} F. Centrone, F. Grosshans, and V. Parigi, Phys. Rev. A 108, 042615 (2023).

\bibitem {ref42} P. Zhu, Y. Wang, Y. Du, M. Yu, K. Zhang, K. Wang, and P. Xu, Sci. China-Phys. Mech. Astron. 68, 260311 (2025).

\bibitem {ref43} S. L. Braunstein, and P. V. Loock, Rev. Mod. Phys. 77, 513 (2005). 

\bibitem {ref44} R. Horodecki, P. Horodecki, M. Horodecki, and K. Horodecki, Rev. Mod. Phys. 81, 865 (2009).

\bibitem {ref45} C. Weedbrook, S. Pirandola, R. Garc{\'i}a-Patr{\'o}n, N. J. Cerf, T. C. Ralph, J. H. Shapiro, and S. Lloyd, Rev. Mod. Phys. 84, 621 (2012).

\bibitem {ref46} W. Asavanant, and A. Furusawa, Phys. Rev. A 109, 040101 (2024).

\bibitem {ref47} S. Yokoyama, R. Ukai, S. C. Armstrong, C. Sornphiphatphong, T. Kaji, S. Suzuki, J. Yoshikawa, H. Yonezawa, N. C. Menicucci, and A, Furusawa, Nat. Photon. 7, 982 (2013).

\bibitem {ref48} W. Asavanant, Y. Shiozawa, S. Yokoyama, B. Charoensombutamon, H. Emura, R. N. Alexander, S. Takeda, J. Yoshikawa, N. C. Menicucci, H. Yonezawa, and A. Furusawa, Science 366, 373 (2019).

\bibitem {ref49} M. Pysher, Y. Miwa, R. Shahrokhshahi, R. Bloomer, and O. Pfister, Phys. Rev. Lett. 107, 030505 (2011).

\bibitem {ref50} M. Chen, N. C. Menicucci, and O. Pfister, Phys. Rev. Lett. 112, 120505 (2014).

\bibitem {ref51} J. Roslund, R. de Ara{\'u}jo, S. Jiang, C. Fabre, and N. Treps, Nat. Photon. 8, 109 (2014).

\bibitem {ref52} D. Wilken, J. Junker, and M. Heurs, Phys. Rev. Applied 21, L031002 (2024).

\bibitem {ref53} X. Jia, C. Zhai, X. Zhu, C. You, Y. Cao, X. Zhang, Y. Zheng, Z. Fu, J. Mao, T. Dai, L. Chang, X. Su, Q. Gong, and J. Wang, Nature 639, 329 (2025).

\bibitem {ref54} E. H. Huntington, and T. C. Ralph, J. Opt. B 4, 123 (2002).

\bibitem {ref55} S. Shi, Y. Wu, L. Gao, L. Zheng, L. Tian, Y. Wang, W. Li, and Y. Zheng, Opt. Lett. 48, 3111 (2002).

\bibitem {ref56} Y. Wu, Q. Wang, L. Tian, X. Zhang, J. Wang, S. Shi, Y. Wang, and Y. Zheng, Photon. Res. 10, 1909 (2022).

\bibitem {ref57} F. Li, X. Zhang, J. Li, J. Wang, S. Shi, L. Tian, Y. Wang, L. Chen, and Y. Zheng, Front. Phys. 18, 42303 (2023).

\bibitem {ref58} C. M. Caves, and P. D. Drummond, Rev. Mod. Phys. 66, 481 (1994).

\bibitem {ref59} F. Meylahn, B. Willke, and H. Vahlbruch, Phys. Rev. Lett. 129, 121103 (2022).

\bibitem {ref60} W. Yao, Q. Wang, L. Tian, R. Li, S. Shi, J. Wang, Y. Wang, and Y. Zheng, Laser Phys. Lett. 18, 015001 (2020).

\bibitem {ref61} H. Vahlbruch, S. Chelkowski, B. Hage, A. Franzen, K. Danzmann, and R. Schnabel, Phys. Rev. Lett. 97, 011101 (2006).

\bibitem {ref62} B. Hage, A. Samblowski, and R. Schnabel, Phys. Rev. A 81, 062301 (2010).

\bibitem {ref63} L. Tian, S. Shi, Y. Li, Y. Wu, W. Li, Y. Wang, Q. Liu, and Y. Zheng, Opt. Lett. 46, 3989 (2021).

\bibitem {ref64} L. Gao, L. Zheng, B. Lu, S. Shi, L. Tian, and Y. Zheng, Light Sci. Appl. 13, 294 (2024).

\bibitem {ref65} C. Jebarathinam, D. Home, and U. Sinha, Phys. Rev. A 101, 022112 (2020).

\bibitem {ref66} Y. Wu, S. Shi, X. Liu, L. Tian, W. Li, Y. Wang, and Y. Zheng, Phys. Rev. Applied 23, 044021 (2025).

\bibitem {ref67} Y. Wu, L. Tian, W. Yao, S. Shi, X. Liu, B. Lu, Y. Wang, and Y. Zheng, Appl. Phys. Lett. 124 (11): 114002 (2024).

\bibitem {ref68} J. Lee, S. Ji, J. Park, and H. Nha, Phys. Rev. A 90, 022301 (2014).

\bibitem {ref69} S. Shi, Y. Wu, L. Gao, L. Zheng, L. Tian, Y. Wang, W. Li, and Y. Zheng, Opt. Lett. 48, 3111 (2023).

\bibitem {ref70} Z. Wang, K. Li, Y. Wang, X. Zhou, Y. Cheng, B. Jing, F. Sun, J. Li, Z. Li, B. Wu, Q. Gong, Q. He, B. Li, and Q. Yang, Light Sci. Appl. 14, 164 (2025).

\bibitem {ref71} S. Liu, Z. Lu, P. Wang, Y. Tian, X. Wang, and Y. Li, npj Quantum Inf. 9, 92 (2023).

\bibitem {ref72} A. Khalique, W. Tittel, and B. C. Sanders, Phys. Rev. A 88, 022336 (2013).

\bibitem {ref73} K. Azuma, S. E. Economou, D. Elkouss, P. Hilaire, L. Jiang, H. K. Lo, and I. Tzitrin, Rev. Mod. Phys. 95, 045006 (2023).

\bibitem {ref74} C. Huang, Y. Chen, T. Luo, W. He, X. Liu, Z. Zhang, and K. Wei, Sci. China-Phys. Mech. Astron. 67, 240312 (2024).

\bibitem {ref75} H. Guo, N. Liu, Z. Li, R. Yang, H. Sun, K. Liu, and J. Gao, Photon. Res. 10, 2828 (2022).

\end{thebibliography}
\end{document}